\documentclass{emulateapj}

\newcommand{\kms}{km\,s$^{-1}$}

\newcommand{\twelveCO}{{}\mbox{$^{12}$}CO}
\newcommand{\thirteenCO}{{}\mbox{$^{13}$}CO}
\newcommand{\jone}{J=1$\rightarrow$0}

\newcommand{\vel}{\rm{v}}

\begin{document}
\title{Discovery Of A Molecular Outflow in the Haro 6-10 Star-Forming Region}
\shorttitle{Molecular Outflows in Haro 6-10}
\shortauthors{Stojimirovi\'c et al.}

\author{Irena Stojimirovi\'c\altaffilmark{1}, Gopal Narayanan, Ronald L. Snell} 
\affil{Astronomy Department, University of Massachusetts at Amherst,
  MA 01003}\altaffiltext{1}{Current address: IAR, Boston University, Boston, MA,02215}

\email{irena@bu.edu, gopal@astro.umass.edu, snell@astro.umass.edu}
\begin{abstract}

We present high sensitivity \twelveCO\ and \thirteenCO\ \jone\
molecular line maps covering the full extent of the parsec scale
Haro~6-10 Herbig-Haro (HH) flow. We report the discovery of a
molecular CO outflow along the axis of parsec-scale HH flow. Previous
molecular studies missed the identification of the outflow probably
due to their smaller mapping area and the confusing spectral features
present towards the object. Our detailed molecular line study of the
full 1.6~pc extent of the optical flow shows evidence for both
blueshifted and redshifted gas set in motion by Haro~6-10
activity. The molecular outflow is centered at Haro~6-10, with
redshifted gas being clumpy and directed towards the northeast, while
blueshifted gas is in the southwest direction. The molecular gas
terminates well within the cloud, short of the most distant HH objects
of the optical flow. Contamination from an unrelated cloud along the
same line of sight prevents a thorough study of the blueshifted
outflow lobe and the mass distribution at the lowest velocities in
both lobes. The cloud core in which Haro~6-10 is embedded is
filamentary and flattened in the east-west direction. The total cloud
mass is calculated from \thirteenCO\ \jone\ to be $\sim 200$~$M_{\odot}$.
The lower limit of the mass associated with the outflow is $\sim$ 
0.25~M$_{\odot}$.
\end{abstract}


\section{Introduction}

In the last ten years, wide-field CCD camera observations have shown
that it is quite common for outflows traced by Herbig-Haro (HH)
objects to attain parsec scale dimensions, at least an order of
magnitude larger than previously thought \citep{rbd97}. It is now
becoming clear that optical HH objects (which are lit up from the
cooling behind shock fronts within the flows), near-infrared H$_2$
jets (tracing shocked regions), and the swept-up CO bipolar molecular
outflows are intimately linked to one another \citep[see for
e.g.][]{bachiller95}. With the development of large-format heterodyne
arrays at millimeter wavelengths and the availability of "on-the-fly"
(OTF) mapping capabilities, millimeter mapping of the full extent of
some of the known parsec-scale HH flows is possible with
increased data sensitivity.
This allows a more thorough study of the observational properties of
the outflows and provides better constraints on the theoretical models
of molecular outflow entrainment mechanism.

\citet{devine99} discovered a giant 1.6~pc long HH flow, centered on
Haro 6-10 at a position angle of 222\arcdeg\ delineated by HH 410 and
HH 411 at its edges, and HH 412 and HH 184E along the HH flow
axis. Haro~6-10 is located in the L~1524 cloud within the B18 region
of the Taurus Molecular Cloud Complex. It is a binary system composed
of an optically visible southern component (Haro~6-10S), and its
infrared companion (Haro~6-10N). High-resolution 3.6~cm VLA map
\citep{reipurth04} strongly suggests that the optical component, Haro
6-10S, is driving the current outflow activity. The same set of data
suggests that there might be an additional stellar companion to Haro
6-10S, more closely bound to it than Haro 6-10N.

There has been other evidence for the outflow activity from the Haro
6-10 system such as 2.12~$\mu$m molecular hydrogen emission, observed
at the position of the northern companion \citep{carr90,
hkl95}. \citet{moma99} found a small HH jet associated with Haro~6-10
itself at a position angle of 195\arcdeg\ with blueshifted radial
velocities. \citet{reipurth04} reported a VLA jet along the same
angle. These latter jet angles are significantly different from the
222\arcdeg\ angle of the parsec scale HH flow. The difference in these
position angles have led some authors to suggest that the outflow from
Haro~6-10 has undergone precession or reorientation 
\citep{reipurth04,devine99}. \citet{devine99} have argued that HH 184F-G 
and HH 184A-B are two different flows originating at the same time from 
two different protostars, probably due to an interaction in the binary pair. 
They have position angles of 162\arcdeg\ and 231-249\arcdeg\ respectively.
HH 184E is the only knot in the HH 184 that is found along the parsec
scale 222\arcdeg\ axis and is the furthest away from driving source. 

In the past, a few attempts have been made to detect molecular CO
outflow from the Haro~6-10 source. Small maps, only a couple of
arcminutes wide, have been made toward Haro~6-10 which presented the
evidence of weak non-Gaussian CO wings emission in the direction of
the optical object \citep{edwards84, levreault88, hog97, hog98,
cbm98}. Hitherto, Haro 6-10 has been considered to be devoid of
molecular outflow \citep{devine99}.

\begin{figure*}[ht]
\hskip -25pt
\epsscale{1.}
\plotone{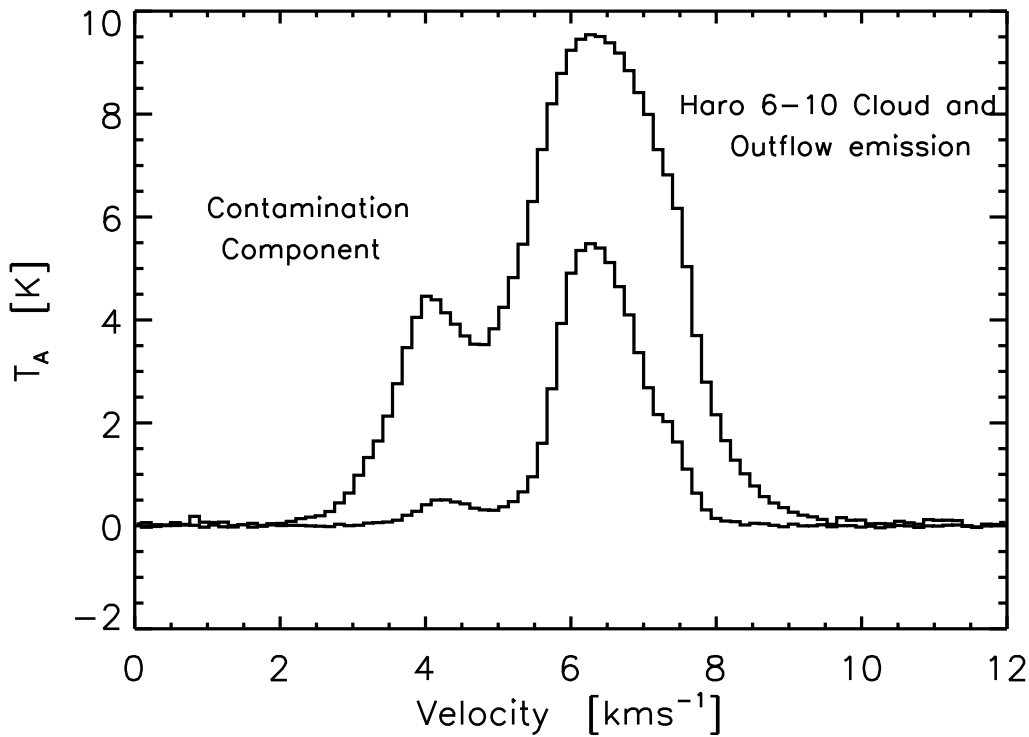}
\caption{The \twelveCO\ and \thirteenCO\ spectra averaged in 15\arcmin\ area 
centered on Haro~6-10. Foreground contamination from another cloud in Taurus 
gives rise to the blueshifted emission peaking at $\sim$ 4.5~\kms. 
\label{fig:av_sp}}
\end{figure*}
\begin{figure*}[ht]
\centering
\epsscale{1.2}
%\plotone{f2.eps}
\caption{(See 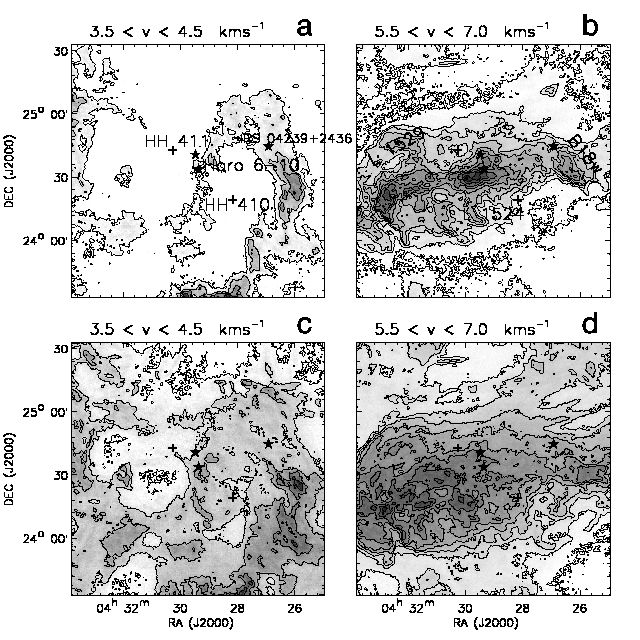) Overview of the Haro 6-10 region. A zoomed view into the 
B18
region derived from a much larger-scale map from the Taurus molecular
cloud survey \citep{taurus06} in \thirteenCO\ \jone\ ({\it a,b}) and
\twelveCO\ \jone\ ({\it c,d}). These images demonstrate the presence of two
emission components at different velocities, along the same line of
sight. Panels {\it a} and {\it c} show emission unrelated to the B~18 cloud, 
while panels {\it b} and {\it d} trace the B~18 cloud sub-structure. In 
panel {\it b} we identify three clouds found in B~18: L~1529, L~1524, B18w, 
from east to west). Several YSOs and T-Tauri stars are embedded in the L~1529
region. L~1524 hosts Haro~6-10 star and nearby HH 414 IRS (filled
stars). B~18w at the western most point, hosts IRAS 04239+2436 source
which drives parsec scale HH 300 flow. In the panel {\it a} we label
Haro~6-10 binary system, the driving source of parsec-scale HH flow
with terminating points at HH 410 toward SW and HH 411 toward
NE. Contours at each panel start at 2$\sigma$ level and go in steps of
4$\sigma$.
\label{fig:B18}}
\end{figure*}

We made large, sensitive, \twelveCO\ \jone\ and \thirteenCO\ \jone\
maps of the Haro 6-10 region, over the full extent of the optical
parsec-scale HH flow.  Here we show the evidence for the redshifted
high velocity CO gas along the outflow axis defined by the HH parsec
scale flow. Due to ``contamination'' by emission from another
molecular cloud along the same line of sight, a thorough study of the
blueshifted outflow lobe is limited.

\section{Observations}

A full mapping of \twelveCO\ and \thirteenCO\ in the \jone\ transition
was performed with the SEQUOIA receiver at the Five College Radio
Astronomy Observatory (FCRAO) 14~m telescope during a period spanning
2003 and 2004. The receiver was configured as a dual-polarized
4$\times$4 array. The orthogonal polarizations of the SEQUOIA array
were averaged to produce spectra with higher signal-to-noise ratio
(S/N). The telescope's half-power beam widths are 45\arcsec\ and
47\arcsec\ for \twelveCO\ and \thirteenCO\ transitions
respectively. Channel maps as well as individual spectra were checked
for any scanning artifacts, baselined and regridded to the
22.5\arcsec\ sampled grid. RMS noise weighting was used to combine the
data.  For all calculations, the antenna temperatures were corrected
for the main beam efficiencies of 0.45 for \twelveCO\ and 0.5 for
\thirteenCO.

Both \twelveCO\ and \thirteenCO\ spectra are smoothed to 0.13~\kms\
channel spacing. The system temperatures (T$_{sys}$) in our
observations range from 400 -- 700 K for \twelveCO\ and between 200 --
500 K for \thirteenCO. Regions mapped with the higher noise level were
repeated, combined and averaged in order to get a constant lower noise
level over the whole extent of the map.  The resulting mean rms per
velocity channel is 0.2~K for \twelveCO\ and 0.1~K for
\thirteenCO. Antenna pointing and focus were checked every few hours
and corrected using SiO masers.

The analysis were done both using the Gildas software package and IDL
software of Research Systems Inc. Detailed studies of the physical
parameters characterizing the outflows were performed using IDL.

\section{Morphology of The CO gas}   

\subsection{Two component emission in B18 cloud}
  
In Figure~\ref{fig:av_sp} we show \twelveCO\ and \thirteenCO\ spectra
averaged over a 15\arcmin\ area centered on Haro~6-10. In both
lines, the averaged spectra consists of two Gaussian-like components: a
stronger component centered at 6.4 \kms\ and a weaker component at
$\sim$ 4~\kms. A similar double peak profile has been observed by
\citet[][hereafter AG]{ag01} by studying the parsec scale outflow from
IRAS 04239+2426 in the adjacent B~18w cloud. They named the component
at lower velocity ``cloud A'', and surmised that is was produced by
another cloud in Taurus along the same line of sight. 

\begin{figure*}[ht]
\centering
\epsscale{1.2}
\plotone{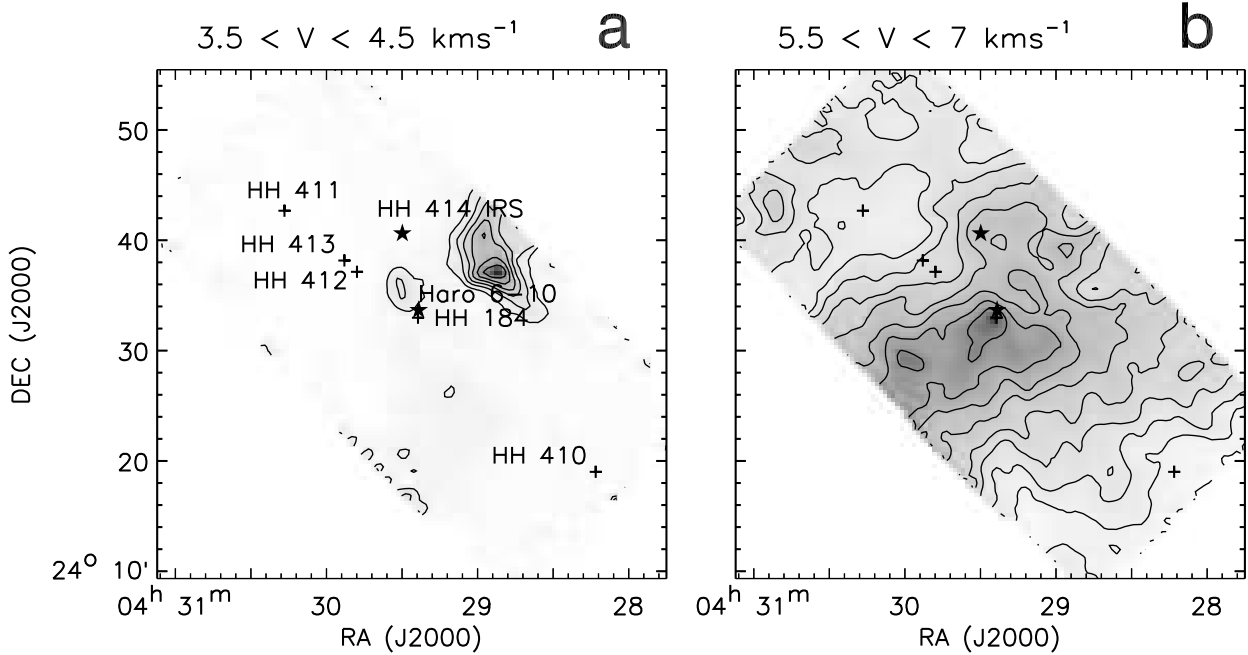}
\caption[L1524 cloud in \thirteenCO\ \jone]
{L~1524 cloud emission using our high sensitivity \thirteenCO\ \jone\
data integrated in two velocity bins, representing two distinct component in
 L~1524 cloud. Contours at each panel start at 5$\sigma$ level and go in steps 
of 10 $\sigma$ in panel {\it a} (0.4~K\kms) and in steps of 20$\sigma$ 
(0.9~K\kms) in panel {\it b}. HH 184 coordinate is taken from Bo Reipurth's 
General Catalog of HH Objects (http://casa.colorado.edu/hhcat/). 
\label{fig:cloud13}}
\end{figure*}

The stronger Gaussian component is associated with the Haro~6-10 host
cloud, L~1524. This is in agreement with the ammonia core identified
at LSR velocity of 6.4~\kms \citep{codella97} towards
L~1524. Haro~6-10 is located at the edge of the compact ammonia core
\citep[see Figure 3, ][]{codella97}.  Our spatially averaged spectra
towards the L~1524 cloud show some evidence for line wing emission at
both blueshifted and redshifted velocities, although the contaminating
cloud component at $\sim 4$~\kms\ obscures much of the wing emission at
blueshifted velocities. Even with our higher sensitivity data, we do
not see evidence for high velocity molecular gas as is seen in other
prototypical outflows. Since the outflow is composed mainly of gas
with slow radial velocities, emission from outflowing gas at the
slowest velocities might be hidden under the ambient cloud emission.

In order to examine this contaminating emission at 4~\kms, we used a
2$^{\circ}$ by 2$^{\circ}$ portion of the much larger FCRAO Taurus
Molecular Cloud Survey in both \twelveCO\ and \thirteenCO\ \jone\
\citep{taurus06}. In Figure~\ref{fig:B18}, we plot for each line, the
integrated intensity emission in two narrow velocity intervals. The
first velocity interval preferentially selects the contaminating
emission in the 3.5 to 4.5~\kms\ velocity range,
(Figure~\ref{fig:B18}{\it a,c}). This contaminating cloud component is
traced by \thirteenCO\ molecule as an arc like filament, stretching
from south of Haro~6-10 to north and appearing to bend toward IRS
04239+2436 in B18w cloud and south from it again, (see
Figure~\ref{fig:B18}{\it a}). The maximum intensity of this emission
is south from IRS 04239+2436 source in the B~18w cloud, while at the
location of Haro~6-10, the contaminating emission is quite clumpy. The
\twelveCO\ tracer shows similar morphology of the emitting gas in the
region (see Figure~\ref{fig:B18}{\it c}).

In the other velocity interval, from 5.5 to 7.0~\kms\ we
preferentially select the CO emission associated with the B~18 clouds,
which have LSR velocities around 6.4~\kms.  In \thirteenCO\ (see
Figure~\ref{fig:B18}{\it b}), the B~18 cloud is revealed as a
flattened, elongated structure in the east-west direction.  In the same
panel we identify L~1529, L~1524 and B~18w clouds within the B~18 region.
 
The dramatic change in the morphology for each isotope at only a
slightly different velocity offset indicates quite clearly that we are
looking at two spatially distinct and possibly unrelated cloud
features.
From Figure~\ref{fig:B18}, it can be seen that towards Haro 6-10, the
intensity of the contaminating emission is not as high as the
intensity of the L~1524 cloud. However, towards B~18w, it is seen that
the emission in both velocity bins seem to be of comparable
intensity. The same was evident in the average spectra of the region
presented by AG. In this study by AG, it was thus a clear cut in the 
case of IRAS 04239+2426 to concentrate only on the redshifted emission 
in the outflow. From the large-scale CO maps derived from the Taurus 
survey, it appears that while the blueshifted component in the 
Haro~6-10 spectra at 4.5~\kms\ is part of the contaminating emission 
seen towards B~18w, the contamination is not nearly as severe as in 
the case of B~18w. Some of the emission towards Haro~6-10 at 4.5~\kms\ 
might be part of the outflow system from that object.
\begin{figure*}[ht]
\epsscale{0.9}%75}
\plotone{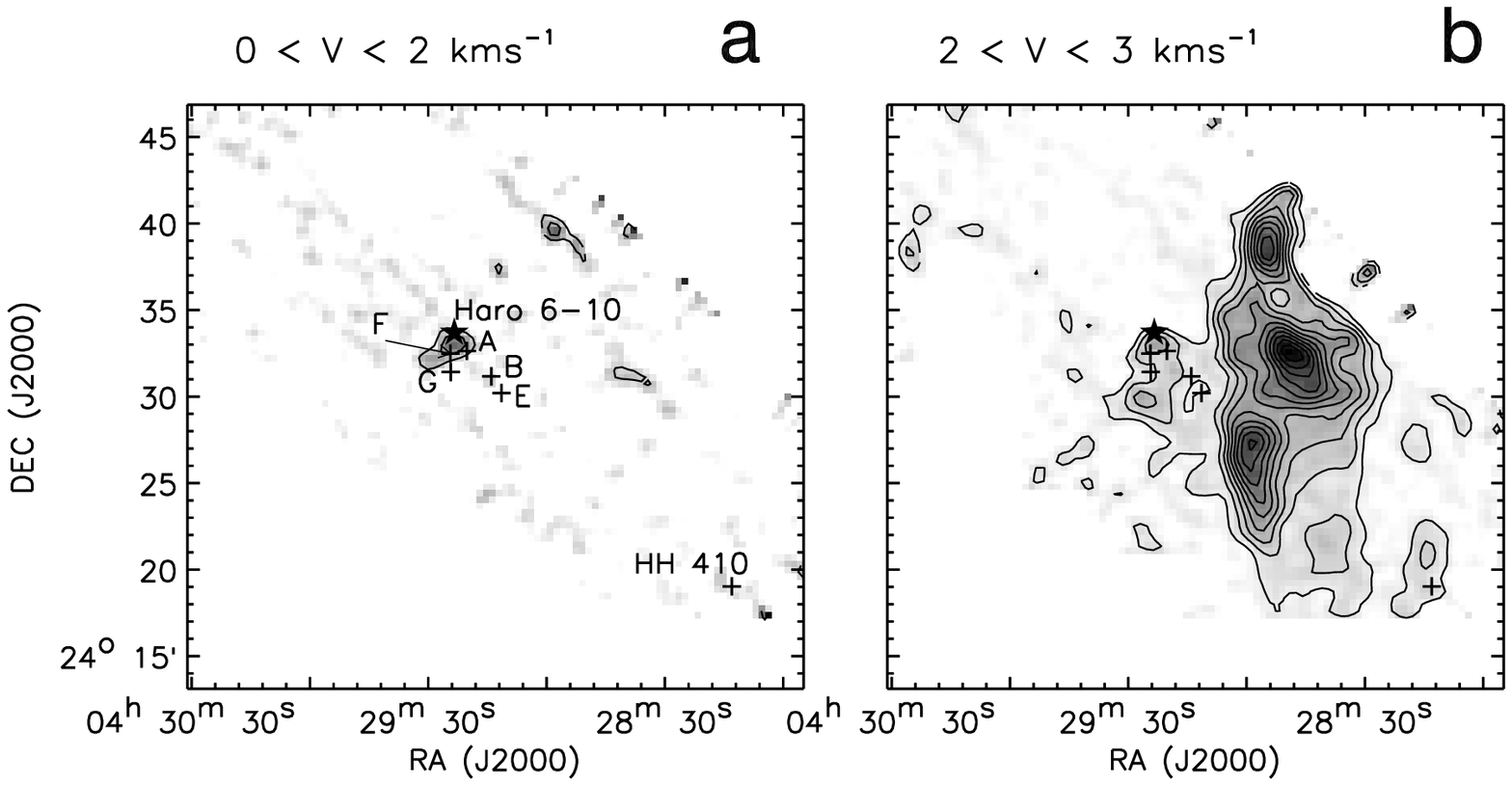}
\vskip-10pt
\epsscale{1.2}%0}
\plotone{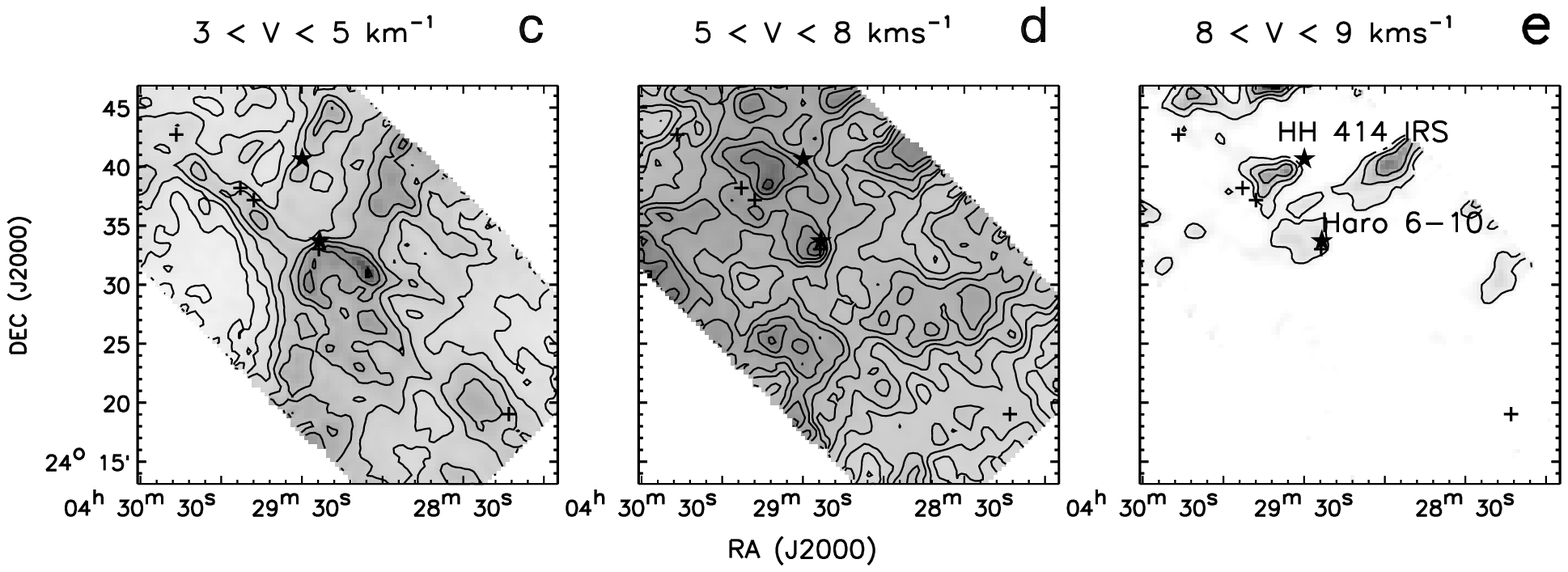}
\vskip-10pt
\epsscale{0.9}%75}
\plotone{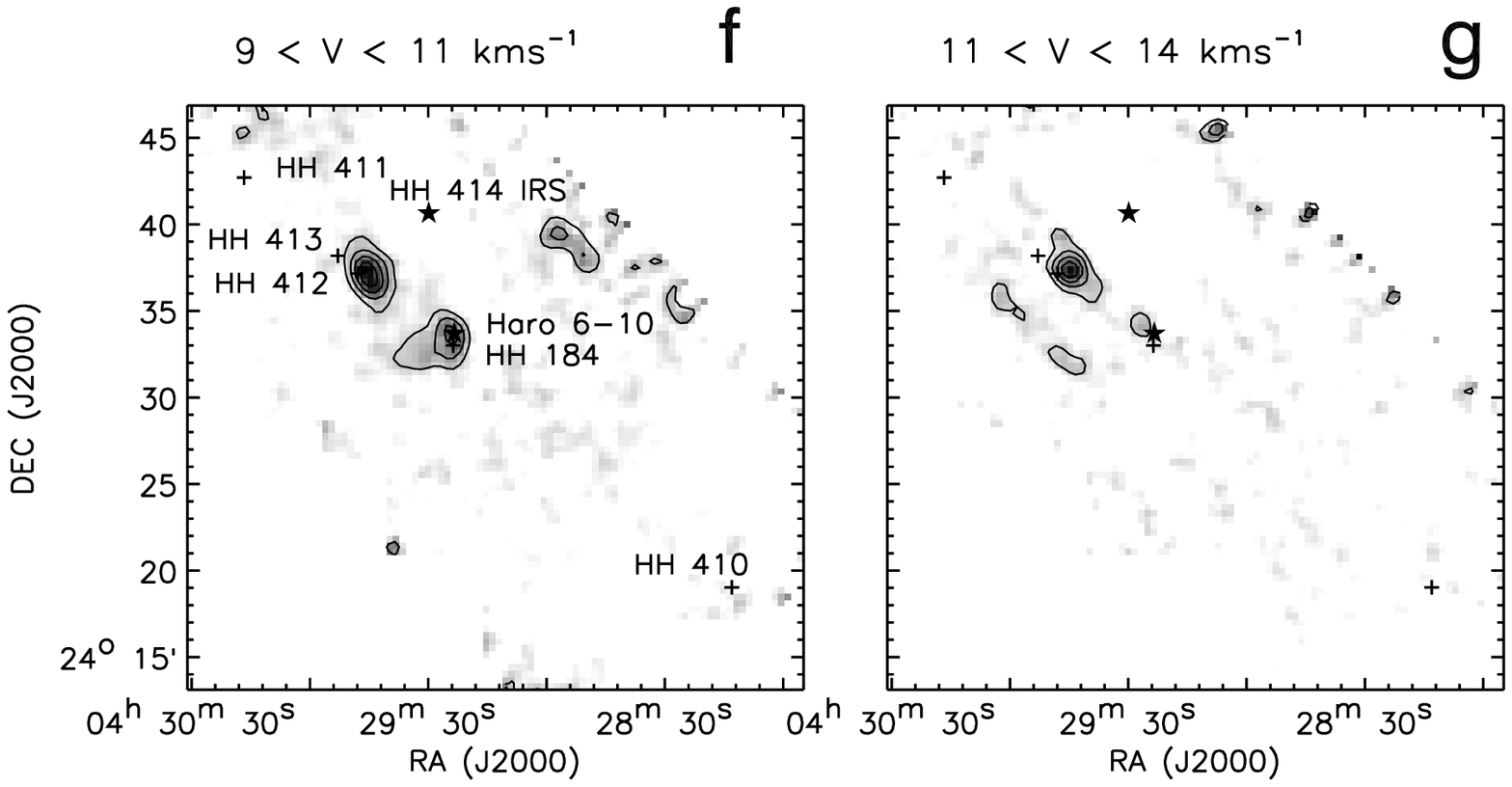}
\caption[Haro~6-10 \twelveCO\ \jone\ channel emission]{\twelveCO\ line 
integrated in different velocity intervals. Upper row: Blueshifted gas. In 
panel {\it a} contours start and increase with $2\sigma$ certainty (2~K\kms), 
while in panel {\it b} they start and increase in 3$\sigma$ steps (2.1~K\kms). 
Middle row: Line Core. Contours in {\it c} and {\it d} panels start at 
10$\sigma$ and go in step of 10 $\sigma$ (1~K\kms\ at panel {\it c} and 
1.2~K\kms\ at panel {\it d}). In panel {\it e} contours start at 8$\sigma$ 
level (0.56~K\kms) and increase in the same step. Lower row: Redshifted gas. At
 both panels contours start and increase with $2\sigma$ certainty, which is 
0.2~K\kms\ in panel {\it f} and 0.32~K\kms\ in panel {\it g}. In {\it a,~b} we 
mark different knots of HH 184 \citep{devine99}, while in the rest of the 
panels we use HH 184 coordinate as defined in Figure~\ref{fig:cloud13}.
\label{fig:mosaic_12}}
\end{figure*}   

\begin{figure*}[ht]
\centering
\epsscale{1.2}
%\plotone{f5.eps}
\caption{(See 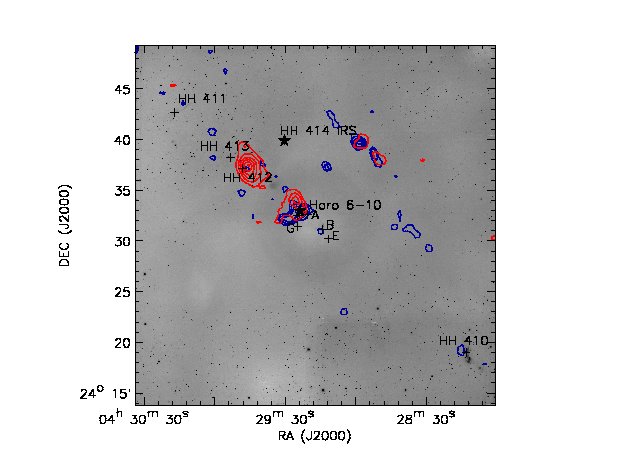) Best case scenario for the outflow in the Haro 6-10 
region:
blueshifted emission is integrated in the velocity interval from 
0 to 2 \kms\ and contours start at 0.3~K\kms\ and 
go in steps of 0.1~K\kms; redshifted emission is integrated in velocity
interval from 9 to 14 \kms\ and contours start at 0.5~K\kms\ 
and go in steps of 0.5~K\kms. Molecular gas is overlayed on the optical
image of the Haro~6-10 region \citep{devine99} Again, we mark A, B, E, G knots 
of the HH 184 source \citep{devine99}. 
\label{fig:outflow}}
\end{figure*}

\subsection{The Haro~6-10 cloud core}

In the following sections, we will examine in more detail the behavior
of the ambient, blueshifted and redshifted gas using our new
\twelveCO\ and \thirteenCO\ \jone\ data of Haro~6-10 outflow, which
are of smaller scale compared to the Taurus survey, but have better
sensitivity. For the rest of this paper, the Taurus Survey data are not used
for any analysis. As already discussed, in the velocity range between
3.5 and 4.5 \kms, the emission is dominated by the contaminating cloud
component. Figure~\ref{fig:cloud13} shows the \thirteenCO\ in two
different velocity bins. Here, the contamination is clearly detected
as strong feature at the northwest edge of the map (see
Figure~\ref{fig:cloud13}{\it a}). In addition, a small clump of
emission centered at Haro~6-10 is present, which is elongated and
extended toward north, at the $\sim$ 30\arcdeg\ angle with respect to
the axis defined by the HH flow (crosses in
Figures~\ref{fig:cloud13}a). In the next panel, \thirteenCO\ emission
traces the column density distribution within the L~1524 cloud. The
flattened structure of the L~1524 cloud seen in the larger-scale map of
Figure~\ref{fig:B18} is also visible here, and Haro~6-10 lies close to
the peak of integrated intensity emission. This peak traces the same
core identified in ammonia by \citet{codella97}. The elongated L~1524
cloud is perpendicular to the HH outflow axis.
From Haro~6-10, along the HH flow axis, the
integrated intensity decreases, and HH 411 and HH 410 are found in the
regions of very low \thirteenCO\ integrated intensity. The ridge of
\thirteenCO\ emission is seen to extend northward from Haro~6-10
towards HH~414 IRS.

\subsection{Outflow Emission}

In Figure~\ref{fig:mosaic_12}, we show integrated intensity emission
of \twelveCO\ in several different velocity bins. 
In the panels {\it~a,~f} and {\it g} we show the distribution of
emission at the highest blueshifted and redshifted velocities.  We
believe that at these velocities we are selecting almost exclusively
emission that arises from the outflow associated with Haro~6-10.
Figure~\ref{fig:mosaic_12}{\it~a} shows that the blueshifted emission in the
velocity range from 0 to 2~\kms\ is spatially restricted and centered
on Haro~6-10. Compared to the bubble-like structures defined by the knots 
A-G of HH 184, the blushifted emission has a smaller spatial extent and is
elongated in the direction defined by A and F knots of HH 184. The spatial 
resolution of our data is such that is does not allow a more detailed 
comparison.
In Figure~\ref{fig:mosaic_12}{\it~f,g} the redshifted
emission consists of two bright regions
centered on Haro~6-10 and HH 412 respectively. The clump
centered at Haro~6-10 has extended emission toward southeast, similar to
the highest velocity blueshifted gas identified in
Figure~\ref{fig:mosaic_12}{\it~a}.
The redshifted emission seen toward HH 412,
located about 7\arcmin\ northeast of Haro~6-10, is detected to
velocities as high as 14~\kms, nearly 8~\kms\ from the ambient cloud 
line center.  The emission is slightly extended along the HH flow
axis.  In the optical images \citep{devine99}, HH 412 appears as a 1\arcmin\
long fairly diffuse emission feature, also elongated along the flow axis.
Therefore, morphologies of the
CO and optical emissions are quite similar around HH 412.

In Figure~\ref{fig:outflow} we show the optical H$\alpha$ image of the
Haro~6-10 region \citep{devine99}, overlayed with the \twelveCO\ \jone\
integrated
emission in 9 to 14~\kms\ velocity range for the redshifted gas and 0
to 2~\kms\ for the blueshifted gas.  In the optical image, HH 410 and
HH 411, lie at the extremes of the HH flow, and are seen against a background
with many galaxies, demonstrating that the flow has bursted out
of the L~1524 cloud in both directions, and is possibly now
moving through a region with greatly
reduced column density.  Our \thirteenCO\ maps with negligible
emission in these regions support this hypothesis.
The expanded image of the CO outflow in Figure~\ref{fig:outflow} shows
that even toward Haro 6-10, the redshifted and blueshifted outflowing
gas are slightly offset, suggesting that the outflow is bipolar with the
redshifted emission to the northeast and blueshifted emission to the
southwest.
\citet{devine99} speculated, based on morphology and brightness, that
the large HH flow has the same bipolar orientation.  Additional
information is provided by the radial velocity measurement of
the small HH jet oriented at position angle 195\arcdeg\  that shows
that it is blueshifted \citep{moma99}, and thus has the same velocity
sense at the CO outflow, although the jet is
oriented about 30\arcdeg\ from the large HH flow.

The distribution of CO emission in the near outflow wings, 2 to 3~\kms\
for the blueshifted gas and 8 to 9~\kms\ for the redshifted gas, is
shown in Figure~\ref{fig:mosaic_12}{\it~b,e}.  
On the blueshifted side, emission is seen toward Haro~6-10 (as in 
Figure~\ref{fig:mosaic_12}{\it~a}) along with much brighter and very extended
 emission oriented in a north-south direction. The southern most part of 
this emission lies on the HH flow axis and may be associated with the flow. 
However, most of the emission presumably has nothing to do with the flow.  
Therefore the emission in the 2 to 3~\kms\ range may be a mix of 
emission from the contaminating gas component as well as emission from the 
outflow. 
It is interesting to note that at these velocities the CO emission close 
to Haro 6-10 is extended in the direction toward knots F and G in HH 184. 
Because of the contamination
in this velocity interval we can not be sure whether this is outflow
related or unrelated emission.
Likewise in the 8 to 9~\kms\ range there is emission near Haro 6-10
that may be related to the CO outflow.  An additional feature is found
to the north-east of Haro 6-10 and connects HH 414 IRS to HH 413 and
HH 412.  \citet{devine99} detected a small jet originating from HH
414 IRS (IRAS 04264+2433) that was directed toward HH 413.  They also
suggested that the morphology of HH 413 knots resembled bow shocks
originating from the direction of HH 414 IRS. The morphology of the CO
emission is suggestive that it is tracing the molecular counterpart 
to the HH 414 IRS flow. 
There could be also some contamination at these 
velocities from the overall Haro 6-10 flow
oriented north-east to south-west. It is noteworthy that the outflow
axis, defined by the HH 414/413 optical flow and by the CO emission
connecting them, is nearly perpendicular to the axis of the main
parsec-scale HH outflow.  Bright CO emission is also seen north-west
of Haro 6-10 that is presumably unrelated to either flow.  Again the
emission in this velocity range is likely a mix of outflow emission
and emission from the ambient gas.

\begin{figure}[ht]
\centering
\epsscale{1.2}
\plotone{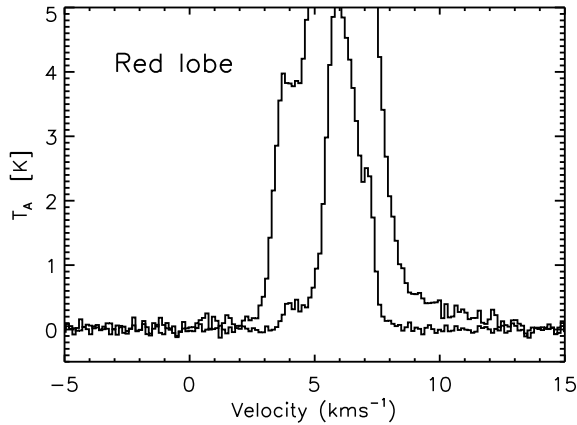}
\caption{\twelveCO\ (thick) and \thirteenCO\ (thin) spectra averaged in the 
redshifted lobe of the Haro~6-10 outflow. 
\label{fig:mv_red}}
\end{figure}

Finally it is interesting to note the structure in the 3 to 5~\kms\ interval,
Figure~\ref{fig:mosaic_12}{\it c}.  This velocity interval would be
expected to include the outflow emission from Haro 6-10, since it lies
a few ~\kms\ from the line center of the L1524 cloud emission, however
the second cloud contaminates the emission. It is intriguing that there
a finger-like emission from Haro~6-10 toward the northeast, and a bow shaped 
feature centered on Haro~6-10, and opening toward HH 410 in the southwest. 
These two features line up with the HH flow direction. 
The relevance of this structure to the outflow is unclear.
In the following velocity bin (5
to 8~\kms), the \twelveCO\ emission is dominated by L~1524 ambient
cloud emission and no clear structure can be identified, although
there is a maximum of emission towards Haro~6-10.

\section{Mass and Energetics}

The emission from \thirteenCO\ is usually optically thin and
therefore a good tracer of the gas column density.  
However, the emission can also be very weak, limiting the 
spatial and velocity extent to which outflows can be traced.
While \twelveCO\ is more readily detected, it is often optically 
thick, even in the high velocity emission of outflows.  To
derive accurate gas column densities we may need to correct for the 
optical depth of this line.  Several recent studies
have combined \twelveCO\ and \thirteenCO\ \jone\ data to estimate the
mass of the outflow, using a velocity-dependent opacity correction
\citep{brlb99, ybb99, ag01, stojimirovic06}.   This approach usually
requires construction of spatially averaged spectra, since \thirteenCO\
emission is usually not detectable in most mapping positions.
The application of this method in Haro~6-10 has some limitations due
to the weak emission in the line wings.
\begin{figure}[ht]
\centering
\epsscale{1.2}
\plotone{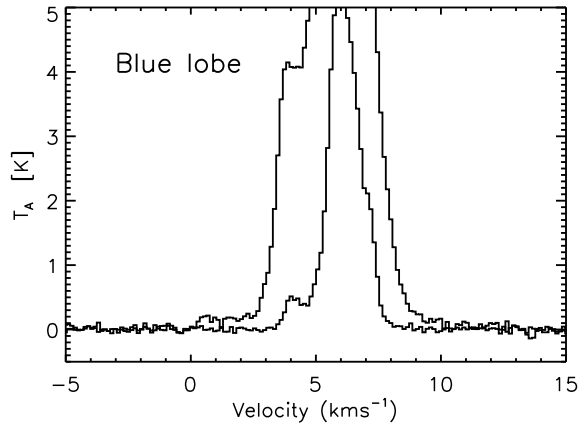}
\caption{\twelveCO\ (thick) and \thirteenCO\ (thin) spectra averaged in the 
blueshifted lobe of the Haro~6-10 outflow. \label{fig:mv_blue}}
\end{figure}
\subsection{Outflow Mass}

In Figures~\ref{fig:mv_red} and \ref{fig:mv_blue} we show the
spatially averaged \twelveCO\ and \thirteenCO\ spectra in the
redshifted and blueshifted gas respectively.  These spectra
were constructed using the following procedure to isolate the
strongest outflow emission. 
For the blueshifted gas we found the integrated intensity at each
location in our \twelveCO\ map in the velocity range from 0 to 2~\kms. 
We then averaged the spectra at every location where the integrated
intensity of \twelveCO\ was at least one-third of the peak integrated
intensity.  The \thirteenCO\ spectrum was obtained by averaging the
same spatial locations as was used to form the \twelveCO\ spectrum.
For the redshifted gas we applied the same approach integrating over
the velocity range of 9 to 12~\kms. 
The resulting spectra differ from the average spectrum shown in 
Figure~\ref{fig:av_sp} in having both a weaker contaminating component 
and enhanced outflow wings.  The outflow in \twelveCO\ can be traced 
over a full velocity range from 0 to 14 ~\kms.

Despite this averaging approach, the \thirteenCO\ emission outside
the line cores, has low signal to noise.
Examining both Figures~\ref{fig:mv_red} and
\ref{fig:mv_blue} we see that at the full spectral resolution,
the \thirteenCO\ emission is lost in noise at velocities below 
than 2~\kms\ and greater than 9~\kms.  The \thirteenCO\ outflow
emission is too weak to use velocity-dependent opacity correction.
We have instead integrated the emission in four velocity intervals.
In the near-wings of the redshifted outflow emission (between 8 and
9~\kms) the ratio of \twelveCO\ to \thirteenCO\ is 112 $\pm$ 20, consistent
with the \twelveCO\ emission being optically thin.  At higher outflow
velocities (between 9 and 12~\kms), \thirteenCO\ emission is not
detected, and the 1$\sigma$ lower limit on this ratio is 47.  
  The \thirteenCO\
emission in the blueshifted gas is even weaker, and only limits
on the ratio can be obtained.  In the near blueshifted outflow
(2 to 3~\kms) the 1 $\sigma$ lower limit is 45 and at higher
blueshifted velocities (0 to 2~\kms) the lower limit is 20.
Thus, where \thirteenCO\ emission is detected it is consistent with
optically thin \twelveCO\ emission and everywhere else we can
only determine an upper limit to the \twelveCO\ optical depth.
Therefore to estimate the gas column density and outflow mass,
we will assume that \twelveCO\ emission is optically thin 
(equation 3 from \citet{stojimirovic06}), and we will apply no 
correction for the optical depth. 

The outflow mass as a function of velocity and position is then 
computed from
$M(\vel)=2\mu m_{H}A N_{H_2}(\vel)$, where $\mu=1.36$ is the mean
atomic weight including He and other constituents,
m$_{H}$ is the mass of the hydrogen atom and A is the physical area of
one pixel at the distance of the source. N$_{H_2}$ is the molecular
hydrogen column density obtained using the relation
N$_{H_2}=1.1\times10^4$N$_{12}$ by \citet{flw82} for the Taurus cloud; this
result is consistent with more recent determinations, summarized by
\citet{hlh04}, for other nearby dark clouds. The greatest uncertainty in 
our mass estimates comes from the uncertainty in the N$_{H_2}$ 
to N$_{12}$ ratio \citep{flw82}.

The gas column density derived assuming that CO is in LTE depends
on the gas temperature.
For a gas temperature of 10~K we find 0.02~M$_{\odot}$ in the 
velocity range from 9 to 14~\kms, while the mass in the 
velocity range from 0 to 2~\kms\ is found to be 0.01~M$_{\odot}$. 
For a gas temperature 
of 25~K, in the same velocity ranges, the redshifted gas has mass of 
0.04~M$_{\odot}$ and blueshifted of 0.02~M$_{\odot}$.  We believe
that these velocity ranges are dominated by outflow emission.

The cutoff velocity of 9~\kms\ for the start of the redshifted
component of the outflow and 2~\kms\ for the start of the blueshifted component
correspond to velocity offsets from the systemic velocity of the cloud
of 2.5 and 4.5~\kms, respectively.  There is likely slower-moving gas at
lower redshifted and blueshifted velocities that is missed in our analysis.
Within the blueshifted velocities, the presence of the contaminating
emission at 4.5~\kms\ clearly makes it problematic to single out the
low-velocity outflow component. Our outflow mass estimates are therefore lower
limits. In an effort to include any lower-velocity outflow gas, we have 
repeated the mass calculation for the redshifted velocity range of 8 to 9~\kms\
 and derive a mass of 0.15 Mo, and for the blueshifted velocity range
of 2 to 3~\kms\ and derive a mass of 0.04~M$_{\odot}$, both assuming a gas
temperature of 25 K. Our mass estimates are summarized in Table~\ref{tbl:mass}.

The mass distribution with velocity for molecular outflow has been shown
to have a power-law dependence, such that $M(\vel) \propto
\vel^{-\gamma}$ \citep{richer2000}.  We have derived the mass-velocity
relation for Haro 6-10 using the \twelveCO\ emission and assuming
it is optically thin as we assumed earlier.  We computed the mass
in each velocity channel within the outflow and plotted this
versus  the velocity offset from the host cloud's mean velocity.
The velocity range
over which outflow gas is detected in Haro 6-10 is very limited,
making the determination of this relation very uncertain.
In a log-log plot, the slope of the linear
fit determines the $\gamma$ index. The value of $\gamma$ for the
blueshifted gas is 6 $\pm$ 0.6 and for the redshifted gas
is 5.5 $\pm$ 0.6.

\begin{deluxetable}{ccccc}
\tabletypesize{\scriptsize}
\tablewidth{0pt}
\tablecaption{Outflow Mass Estimates\label{tbl:mass}} \tablewidth{0pt}
\tablehead{ \colhead{} &\colhead{} & \colhead{T$_{\rm ex}$ = 10 K} & \colhead{} & \colhead{T$_{\rm ex}$ = 25 K} \\
\colhead{} &\colhead{} &\colhead{Mass} & \colhead{} &\colhead{Mass}\\
\colhead{Velocity Interval}& \colhead{} &\colhead{M$_{\odot}$}  &\colhead{} & \colhead{M$_{\odot}$}}
\startdata
Red 8 to 9~\kms & \colhead{} &0.1 & \colhead{} &0.15\\
Red 9 to 14~\kms& \colhead{} &0.02 & \colhead{} &0.03 \\
Blue 0 to 2~\kms&\colhead{} & 0.01 & \colhead{} & 0.02  \\
Blue 2 to 3~\kms&\colhead{} &0.02 & \colhead{} &0.04  \\
\tableline
Red Total & \colhead{} &0.12 & \colhead{} & 0.18  \\
Blue Total & \colhead{} & 0.03 & \colhead{} & 0.06 \\
\enddata
\end{deluxetable}

\subsection{Cloud Mass and Energy} 

We determine the cloud mass by using the \thirteenCO\ map,
Figure~\ref{fig:cloud13}{\it b}. The line
center optical depth at each point in the map is derived from the
\thirteenCO\ peak temperature. The excitation temperature is obtained
at each position by solving the radiative transfer equation for the
excitation temperature, assuming \twelveCO\ line to be optically
thick. We find, the total mass of the cloud using this method to be
$\sim 200$~M$_{\odot}$. The cloud mass estimate is mostly free from
the foreground contamination, since we only searched for peak
\thirteenCO\ emission in the narrow velocity range around line center.

The kinetic energy of the cloud is dominated by the turbulent energy of the 
cloud, and is estimated using 
E$_{\rm turb} = 3/(16ln2)M_{\rm cloud}\Delta \vel^2 $. 
The mean turbulent velocity of the ambient gas is determined from the 
\thirteenCO\ line profiles in the cloud. We find the full 
line width at the half maximum to be $\Delta \vel =$ 1.2~\kms. We find that 
kinetic energy of the L~1524 
cloud is $\sim 1.6 \times10^{45}$~ergs.  

\section{Discussion}

Although at the lowest outflow velocities, the blueshifted emission is
strongly contaminated by the foreground cloud emission, we see
signatures of the outflow from Haro~6-10 at the higher blueshifted
velocities.  There is significant overlap between blueshifted and
redshifted emission at the position of Haro~6-10, which together with
relatively low velocities observed in the line wings, suggests that
the flow is in the plane of the sky.  If the flow is in the plane of
sky, slower radial velocities are hidden under the ambient cloud
emission. In such a scenario, the clumps that we see defined in the
redshifted gas (see for instance Figure~\ref{fig:outflow}) are probably
the highest velocity components seen projected in the map.

\begin{figure}[!ht]
\centering
\epsscale{1.2}
\plotone{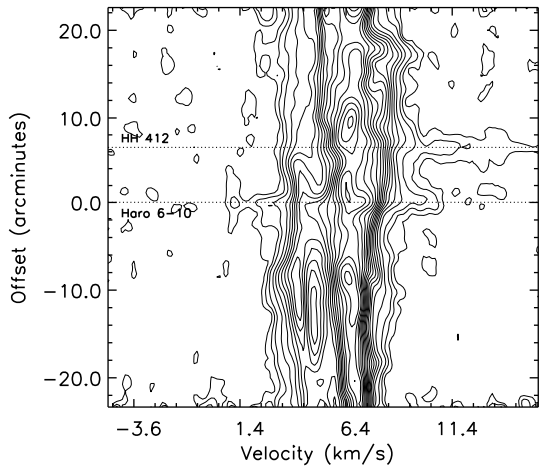}
\caption{Position-Velocity cut for the Haro~6-10 outflow, along the axis 
defined by the parsec-scale HH flow, position angle 40\arcdeg.  Haro~6-10 is
located at an offset position of 0, and HH~412 is located at offset position
of +6.3\arcmin. The cut is wide 5 pixels, which corresponds to $\sim$2\arcmin.  
\label{fig:pv}}
\end{figure}

A position-velocity (P-V) cut along the axis of the HH flow is shown in
Figure~\ref{fig:pv}.  High velocity emission is seen at the location
of Haro~6-10 (at offset 0 in the P-V plot) and at the position of
HH~412 (at offset 6.3).  The highest velocity red-shifted emission
is associated with HH~412.  
The P-V plot also reveals a general broadening of the CO emission at
redshifted velocities between offsets of 0 and $+20\arcmin$, and at
blueshifted velocities between offsets of 0 and $-10\arcmin$. The
broadening at these positions is not symmetric about the cloud's
systemic velocity, but rather, is wider at redshifted velocities for
positive offsets and at blueshifted velocities for negative offsets.
The positive and negative positional offsets in Figure~\ref{fig:pv}
correspond to the location of redshifted and blueshifted lobes
respectively of the overall Haro 6-10 outflow system.  The slight
broadening at positive and negative offsets in the PV plot further
bolsters the argument that the Haro~6-10 molecular outflow is, at
least in the current epoch, a very low-velocity outflow that has
slowed down close to ambient cloud velocities. If this is correct,
it may be possible to see the subtle signature of the
outflow at low velocities from constructing a centroid velocity map
with velocities confined to the ambient cloud emission alone (6 to 7~\kms). 
When such centroid velocity mapping
was performed, for instance in the Cepheus~A outflow system
\citep[see Figure 4 of][]{narayanan96}, it was found that the
outflow signature is clearly seen as a velocity gradient along the
outflow direction. We have performed such centroid velocity mapping
analysis (figure not shown) for Haro~6-10, but we do not see any such
gradient present at ambient cloud velocities. This leads us to
conclude that even if the outflow has slowed to ambient cloud
velocities in Haro~6-10, the outflow has not imparted any significant
velocity gradients over the normally present random velocities
inherent in the ambient cloud.

Yet another puzzling feature of the Haro~6-10 outflow is, that it
entrains very little gas. The morphology of the outflow is very 
clumpy and the total mass found in the outflow in only 0.25~M$_{\odot}$. In 
evolved sources, the previous outflow episodes may clear up the ambient gas 
and leave no gas to be entrained by later outflow episodes.
In such a case, large ambient gas 
cavities are expected to be seen around source. Our \thirteenCO\ maps 
show that no such cavities are found around Haro 6-10 and that the 
overall L~1524 cloud is somewhat flattened and filamentary, with the 
column density decreasing from Haro 6-10 source toward HH 410 and HH 411 
along the outflow axis. While the L~1524 cloud is flattened, there appears 
to be enough ambient gas in the direction of the outflow lobes to be entrained 
(see Figure~\ref{fig:cloud13}). 

The clumpy morphology and low total mass may be a consequence of the outflow 
viewing geometry. If the outflow is mostly in the plane of sky, much of  
its mass 
would be missed since we have excluded ambient gas components from our 
calculations. It would be interesting 
to obtain proper motion study of the HH knots to determine the orientation of
the flow with respect to the plane of the sky. 

The outflow emission beyond Haro 6-10 is strictly confined to the HH 412 knot
 and it has the same elongated morphology as its optical counterpart. 
Therefore
 they seem to be created by the same mechanism. The HH knots 
are formed where the fast jet is interacting with the ambient gas. They can 
be 
found at the head of the jet or along its body at the locations of internal 
working surfaces where the new ejecta is catching up with the previously 
ejected slow-down gas. The CO clump at HH 412 is detached from the 
Haro~6-10 source. It has small mass, and it is detected up to the projected 
velocities of 8~\kms\ from the line center.       

Due to various contaminations, we had to use velocities far from the
line core to study the mass distribution in the case of
Haro~6-10. This could be another reason why there is so little mass
recovered in the outflow. Observational studies show that outflows
show a power law distribution of the mass with respect to the velocity
offset from the cloud velocity. This distribution is such that most of
the outflow mass is at low velocities. The first mass point that we
account for, in Haro~6-10 outflow, is offset by 2.6~\kms\ from the
line center. For the V$_{LSR}$ of 6.4~\kms\, we calculated the outflow
mass only for the points between 9-14~\kms\ and 0 to 2~\kms. Therefore we 
may be missing a significant fraction of the mass as it is increasing 
with the decreasing outflow velocity, and the estimated outflow masses 
are consequently lower limit. 

Lastly, we can compare Haro~6-10 outflow to the HH 300 parsec scale 
flow in the nearby B~18w cloud, studied by AG. AG found 
that the HH 300 outflow has a very clumpy structure and they identified five 
\twelveCO\ redshifted clumps with masses from 0.03 to 0.23~M$_{\odot}$ and 
radial velocities of about 3~\kms\ from the ambient cloud velocity. They also 
noted that the clumps have different position angles with respect to the 
driving source and concluded that the flow most likely precess with each clump 
corresponding to the different ejection event. 
The similarity of the HH 300 and Haro 6-10 outflow in terms of clumpy 
morphology and mass content is apparent. However while the redshifted clumps of
 HH 300 flow are found at different position angles with respect to the IRAS 
source the redshifted clump at HH 412 location in Haro 6-10 flow is found 
elongated along the jet (as is HH 412) axis defined by HH flow. 
Although there are evidence that the current jet 
axis from the Haro 6-10 differs from the axis defined by the parsec scale 
flow \citep{reipurth04, devine99, moma99}, the jet in Haro 6-10 must have had a
 stable orientation for a long time since all HH knots in the Haro 6-10 outflow
 which are found at large distance from the source (HH 410, 411, \& 412) as 
well as the HH 184E, lie on the 222\arcdeg\ position angle jet axis. 
The HH 300 outflow exhibits larger impact on 
its host cloud, with \thirteenCO\ line showing bipolar structure of the cloud 
gas at velocities close to the line center. Surprisingly, Haro 6-10 outflow 
does not seem to be affecting its host cloud's kinematics significantly.  

AG find a steep broken power law in the mass-velocity distribution 
with $\gamma$ = 4.0 at low outflow velocities and $\gamma$ = 7.8 for outflow 
velocities greater than 1.85~\kms. 
We find the slope $\gamma$ in the mass-velocity distribution of the
Haro~6-10 outflows to be around $\sim 6$. This value is much steeper than
the average of $\gamma \sim 2$ reported for a collection of outflow
sources \citep{richer2000}. The steepening of the $\gamma$ index is expected 
for the older outflows, where once accelerated ambient material will slow down 
leading to accumulation of slow material. It is worth noting that the break in 
the mass-velocity distribution in the nearby HH 300 outflow happens very close 
to the line core. If the same holds for the Haro 6-10 we may be missing the 
shallower part at low velocities since we only account for the mass at 
velocities which have offset of at least 2.5~\kms\ (toward redshifted) and 
4.5~\kms\ (toward blueshifted) from the systemic velocity of the cloud. 
The broken power law is usually a good indicator of the jet
entrainment model \citep{zz97}.

\section{Summary And Conclusions}

We made large, sensitive, \twelveCO\ \jone\ and \thirteenCO\ \jone\ maps of 
the Haro 6-10 region, over the full extent of the optical parsec-scale HH flow.
Here we summarize our main conclusions:

\begin{enumerate}
\item
The redshifted outflow component is clearly detected emerging from
Haro~6-10 toward northeast, along the axis of optically defined
parsec-scale HH flow.
\item
Contamination from an unrelated foreground cloud along the same line of 
sight prevents a thorough study of the blueshifted outflow lobe, which we 
detect centered on Haro~6-10 in the opposite direction from redshifted lobe, 
and in the velocity range from 0 to 2~\kms. 
\item
The mass and energies of the outflow are significantly smaller then those of the 
host molecular cloud. However, most of the mass might be missed due to an 
unrelated contamination at lower outflow velocities, along the same line of 
sight, and/or in the plane of the sky orientation of the outflow.  
\item
In the optical maps next to the HH 412, Haro~6-10 parsec scale flow is crossed 
by HH 414/413 flow. In CO data at the lowest redshifted velocities, there seem 
to be evidence of HH~414 IRS activity toward HH~413.

\end{enumerate}
\acknowledgments 
This work was supported by NSF grant AST 02-28993
to the Five College Radio Astronomy Observatory.

Facilities: \facility{FCRAO}


\begin{thebibliography}{}

\bibitem[Arce \& Goodman(2001)]{ag01} Arce, H.~G.~\&
Goodman, A.~A.\ 2001a, \apj, 554, 132
%HH 300

\bibitem[Bachiller et al.(1995)]{bachiller95} Bachiller, R., 
Guilloteau, S., Dutrey, A., Planesas, P., \& Martin-Pintado, J.\ 1995, 
\aap, 299, 857 

\bibitem[Bally et al.(1999)]{brlb99}
Bally, J., Reipurth, B., Lada, C.~J., \& Billawala, Y.\ 1999, \aj, 117, 410

\bibitem[Chandler et al.(1998)]{cbm98} Chandler, C.~J., 
Barsony, M., \& Moore, T.~J.~T.\ 1998, \mnras, 299, 789 

\bibitem[Carr(1990)]{carr90} Carr, J.~S.\ 1990, \aj, 100, 1244 

\bibitem[Codella et al.(1997)]{codella97} Codella, C., Welser, 
R., Henkel, C., Benson, P.~J., \& Myers, P.~C.\ 1997, \aap, 324, 203 

\bibitem[Devine et al.(1999)]{devine99} Devine, D., Reipurth, 
B., Bally, J., \& Balonek, T.~J.\ 1999, \aj, 117, 2931 

\bibitem[Edwards \& Snell(1984)]{edwards84} Edwards, S., \& 
Snell, R.~L.\ 1984, \apj, 281, 237 

\bibitem[Frerking et al.(1982)]{flw82} Frerking,
M.~A., Langer, W.~D., \& Wilson, R.~W.\ 1982, \apj, 262, 590

\bibitem[Harjunp{\"a}{\"a} et al.(2004)]{hlh04}
Harjunp{\"a}{\"a}, P., Lehtinen, K., \& Haikala, L.~K.\ 2004, \aap, 421,
1087

\bibitem[Herbst et al.(1995)]{hkl95} Herbst, T.~M., Koresko, 
C.~D., \& Leinert, C.\ 1995, \apjl, 444, L93 

\bibitem[Heyer et al.(1987)]{heyer87} Heyer, M.~H., Snell, 
R.~L., Goldsmith, P.~F., \& Myers, P.~C.\ 1987, \apj, 321, 370 

\bibitem[Hogerheijde et al.(1997)]{hog97} Hogerheijde, M.~R., 
van Dishoeck, E.~F., Blake, G.~A., \& van Langevelde, H.~J.\ 1997, \apj, 
489, 293 

\bibitem[Hogerheijde et al.(1998)]{hog98} Hogerheijde, M.~R., 
van Dishoeck, E.~F., Blake, G.~A., \& van Langevelde, H.~J.\ 1998, \apj, 
502, 315 

\bibitem[Langer \& Penzias(1993)]{lp93} Langer, W.~D.~\&
Penzias, A.~A.\ 1993, \apj, 408, 539

\bibitem[Lee et al.(2000)]{lee00} Lee, C., Mundy, L.~G.,
Reipurth, B., Ostriker, E.~C., \& Stone, J.~M.\ 2000, \apj, 542, 925

\bibitem[Lee et al.(2001)]{lee01} Lee, C.-F., Stone, J.~M.,
Ostriker, E.~C., \& Mundy, L.~G.\ 2001, \apj, 557, 429

\bibitem[Lee et al.(2002)]{lee02} Lee, C.-F., Mundy, L.~G.,
Stone, J.~M., \& Ostriker, E.~C.\ 2002, \apj, 576, 294

\bibitem[Levreault(1988)]{levreault88} Levreault, R.~M.\ 1988, 
\apj, 330, 897

\bibitem[Masson \& Chernin(1993)]{masson93} Masson, C.~R., \&
Chernin, L.~M.\ 1993, \apj, 414, 230

\bibitem[Matzner \& McKee(1999)]{mm99} Matzner, C.~D.~\&
McKee, C.~F.\ 1999, \apjl, 526, L109

\bibitem[Movsessian \& Magakian(1999)]{moma99} Movsessian, 
T.~A., \& Magakian, T.~Y.\ 1999, \aap, 347, 266 

\bibitem[Narayanan \& Walker(1996)]{narayanan96} Narayanan, G., \& 
Walker, C.~K.\ 1996, \apj, 466, 844

\bibitem[Narayanan et al.(2007)]{taurus06} Narayanan, G., Heyer, M. H., Brunt, C.,
Goldsmith, P. F., Snell, R., Tang, Y., and Li, D., 2007, in preparation

\bibitem[Reipurth et al.(1997)]{rbd97} Reipurth,
B., Bally, J., \& Devine, D.\ 1997, \aj, 114, 2708

\bibitem[Reipurth et al.(2004)]{reipurth04} Reipurth, B., 
Rodr{\'{\i}}guez, L.~F., Anglada, G., \& Bally, J.\ 2004, \aj, 127, 1736 
%VLA jets

\bibitem[Richer et al.(2000)]{richer2000} Richer, J.~S., Shepherd,
D.~S., Cabrit, S., Bachiller, R., \& Churchwell, E.\ 2000, Protostars
and Planets IV, 867

\bibitem[Shu et al.(2000)]{shu00} Shu,
F.~H., Laughlin, G., Lizano, S., \& Galli, D.\ 2000, \apj, 535, 190

\bibitem[Stahler(1994)]{stahler94} Stahler, S., 1994, ApJ, 422, 616

\bibitem[Stojimirovi{\'c} et al.(2006)]{stojimirovic06} 
Stojimirovi{\'c}, I., Narayanan, G., Snell, R.~L., \& Bally, J.\ 2006, 
\apj, 649, 280 

\bibitem[Yu et al.(1999)]{ybb99} Yu, K., Billawala, Y., \&
Bally, J.\ 1999, \aj, 118, 2940

\bibitem[Yu et al.(2000)]{yu2000} Yu, K.~C., Billawala, Y.,
Smith, M.~D., Bally, J., \& Butner, H.~M.\ 2000, \aj, 120, 1974

\bibitem[Zhang \& Zheng(1997)]{zz97} Zhang, Q.~\& Zheng, X.\ 1997, \apj, 474,
719

\end{thebibliography}
\end{document}